\documentclass[manuscript]{aastex}

\usepackage{rotating}
\usepackage{tikz}
\usepackage{amsmath}

\slugcomment{Revised 2020 October 03}

\shorttitle{Hilda Asteroid P/2010 H2}
\shortauthors{Jewitt and Kim}

\begin{document}

\title{Outbursting Quasi-Hilda Asteroid P/2010 H2 (Vales)}


\author{David Jewitt$^{1,2}$
and
Yoonyoung Kim$^3$}

\affil{$^1$Department of Earth, Planetary and Space Sciences,
UCLA, 595 Charles Young Drive East, Los Angeles, CA 90095-1567\\
$^2$Department of Physics and Astronomy, University of California at Los Angeles, 430 Portola Plaza, Box 951547, Los Angeles, CA 90095-1547\\
$^3$ Max Planck Institute for Solar System Research, Justus-von-Liebig-Weg 3, 37077 G\"ottingen, Germany \\
}

\email{jewitt@ucla.edu}

\begin{abstract}
Quasi-Hilda asteroid P/2010 H2 (Vales) underwent a spectacular photometric outburst by $\ge$7.5 magnitudes (factor of $\ge10^3$) in 2010.  Here, we present our optical observations of this event in the four month period from April 20 to August 10.  The outburst, starting UT 2010 April 15.76, released dust particles of total cross-section 17,600 km$^2$ (albedo 0.1 assumed) and mass $\sim 1.2\times10^9$ kg, this being about 10$^{-4}$ of the mass of the nucleus, taken as a sphere of radius 1.5 km and density 500 kg m$^{-3}$.  While the rising phase of the outburst was very steep (brightness doubling time of hours),  subsequent fading occurred slowly (fading timescales increasing from weeks to months), as large, low velocity particles drifted away from the nucleus.  A simple model of the fading lightcurve indicates that the ejected particles occupied a broad range of sizes, from $\sim$ 1 $\mu$m to 6 cm, and followed a differential power-law distribution with index 3.6$\pm$0.1 (similar to that in other comets). The fastest particles had speeds $\ge$210 m s$^{-1}$, indicating gas-drag acceleration of small grains well-coupled to the flow.   Low energy processes known to drive mass loss in active asteroids, including rotational disruption, thermal and desiccation stress cracking, and electrostatic repulsion, cannot generate the high particles speeds measured in P/Vales, and are discounted.  Impact origin is unlikely given the short dynamical lifetimes of the quasi-Hildas and the low collision probabilities of these objects.  The specific energy of the ejecta is estimated at 220 J kg$^{-1}$.  The outburst follows a series of encounters with Jupiter in the previous century, consistent with the delayed activation of buried supervolatiles (and/or the crystallization of sub-surface amorphous ice) by conducted heat following an inward displacement of the perihelion.   A potential origin in the debris cloud produced by avalanche is also considered.

\end{abstract}


\section{INTRODUCTION}
Object P/2010 H2 (Vales) was discovered by Jan Vales as a $m_V \sim$ 12.5 magnitude source in data taken with the 0.6 m telescope at Crni Vrh Observatory (Slovenia) on UT 2010 April 16.01 (Vales et al.~2010).  While first noticed as a point source, visual observers began within a day  to report an expanding coma (Vales et al.~2010, Mikuz et al.~2010).  Prediscovery observations include an upper  limit to the brightness  $m_V \gtrsim$ 20 on UT April 15.4 (Catalina Sky Survey observation by R. Kowalski, reported in Vales et al.~2010), another non-detection at unfiltered magnitude $>$16  on April 15.56 (Balanutsa et al.~2010) and a pre-discovery detection at $m_V$ = 13.7$\pm$0.1 on April 15.82 (Balanutsa et al.~2010), the latter two measurements with 0.4 m diameter telescopes of the MASTER network. Combined, the observations show sudden brightening of this previously unknown object with a doubling time of hours.  Outbursts in comets are not rare (e.g.~13 were tabulated in a 9 year period by Ishiguro et al.~2016), but few have amplitudes as large as the $\ge$ 7.5 magnitude range (a factor of $\ge 10^3$) in P/Vales, and few have been observationally well-characterized, providing a motivation for this study.   

\begin{figure}
\epsscale{1.0}
\plotone{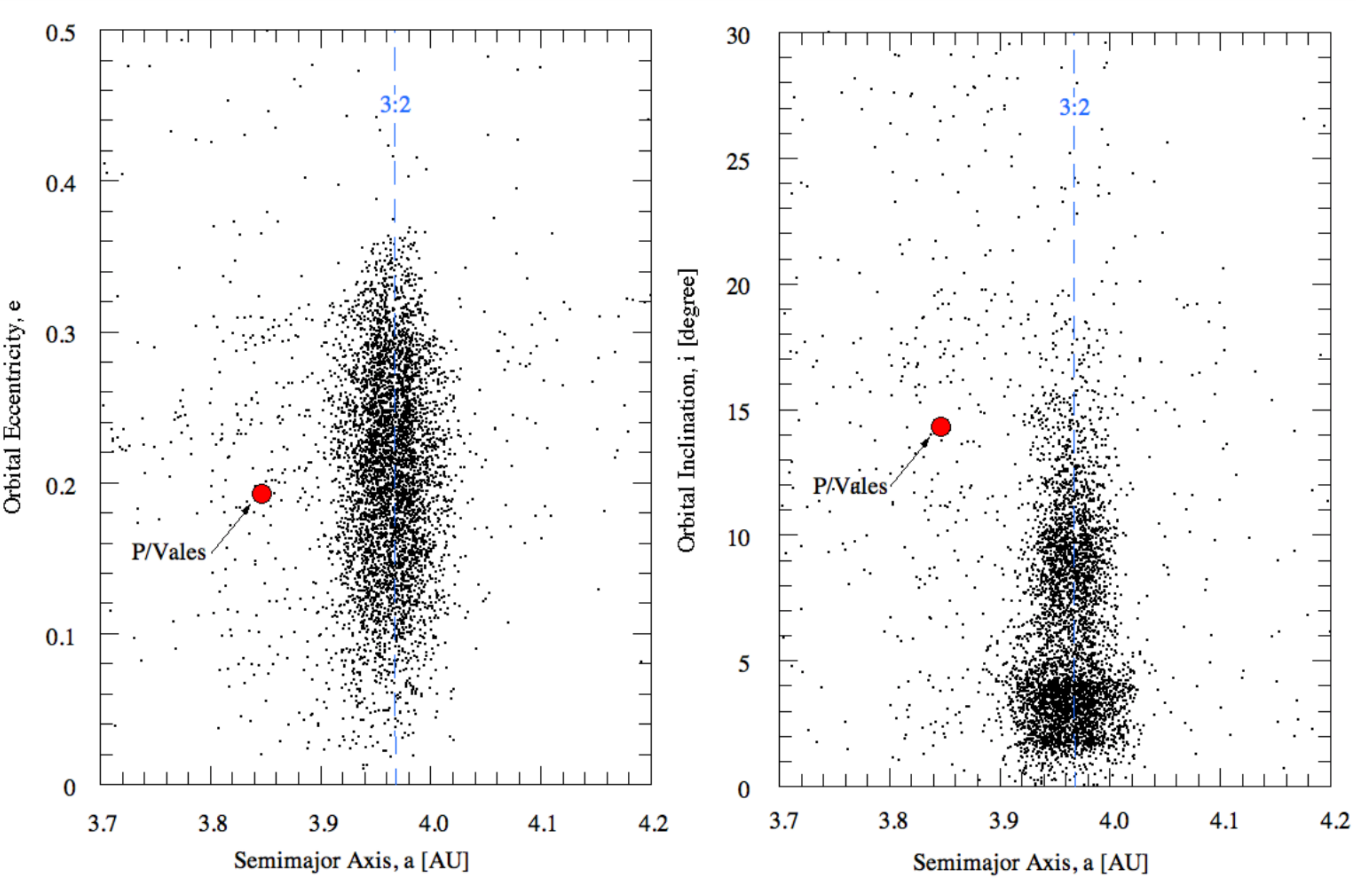}
\caption{(left) Semimajor axis vs.~eccentricity for asteroids near the 3:2 mean motion resonance with Jupiter (marked as a vertical dashed blue line).  The location of P/2010 H2 (Vales) is marked with a red circle. (right)  Same but for semimajor axis vs.~inclination. Obvious bimodal structure in the $a$ vs.~$i$ plot reflects the existence of the Hilda and Schubart collisional families (Vinogradova 2015).   
\label{aei}}
\end{figure}

The orbit has semimajor axis, $a$ = 3.850 AU, eccentricity, $e$ = 0.193 and inclination, $i$ = 14.3\degr, an orbital period of 7.56 year and P/Vales passed perihelion (at 3.108 AU) on UT 2010 March 9.3.    The orbital elements give a Tisserand parameter with respect to Jupiter, $T_J$ = 2.99, signifying strong gravitational interactions with that planet.  This Tisserand is too small to qualify P/Vales as an active asteroid (Jewitt et al.~2015), for which $T_J \ge$ 3.08 is a minimum requirement.  P/Vales has instead been described as a  Hilda asteroid (Hildas are bodies in orbit near the location of the 3:2 mean-motion resonance with Jupiter, at semimajor axis $a$ = 3.971 AU).   P/Vales indeed lies within the  orbital element range of the Hilda group as defined  by Zellner et al.~(1985), namely 3.7 $\le a \le$ 4.2 AU, $e \le$ 0.3, and $i \le$ 20\degr.   However,  its  semimajor axis is smaller than 96\%, and its inclination larger than 92\%, of the $\sim$5000  objects in this semimajor axis range (Figure \ref{aei}).   Even though we refer to the osculating orbital elements (which change with time owing to perturbations from the planets), the figure serves to show that P/Vales is dynamically distinct from the typical Hilda asteroids clustered around the resonance.  Supporting this difference, Marsden (2010) found that P/Vales passed within 1 AU of Jupiter in 1976, and six comparably close encounters have occurred within the 20th century\footnote{JPL Small Body Database Browser: \url{https://tinyurl.com/y3kfyahe}}.  Consequently, we regard P/Vales as a likely ``Quasi-Hilda", presumably a temporarily captured Jupiter-family comet, several examples of which have been previously described in the literature (e.g.~Toth 2006, Gil-Hutton and Garcia-Migani 2016).

The outburst of P/Vales attracted considerable observational attention  over the spring and summer of 2010. We searched for additional detections of P/Vales  from the prior and subsequent orbits in archival data, finding none. While some of the  observations from 2010 were obtained by eye and are of limited scientific value, even  calibrated observations taken with electronic detectors  have failed to find their way into the refereed literature, a situation we begin to reverse with this paper.  

\section{OBSERVATIONS}
Early epoch observations were taken at our request only 4 days after the discovery, on UT 2010 April 20, using the ``Baade'' 6.5 m diameter Magellan telescope, located in Chile, by Scott Sheppard. We used a 2048$\times$4096 pixel section of the IMACS short camera, giving a field of view approximately 400\arcsec$\times$800\arcsec~at the 0.2\arcsec~pixel$^{-1}$ image scale.  The seeing was 0.7\arcsec~to 0.8\arcsec~Full Width at Half Maximum (FWHM).    A total of 24 images of P/Vales were secured, with a range of exposure times from 2 s to 300 s.  After rejecting images in which the core of the target was saturated, we analysed 3 images in Sloan g', 9 in Sloan r', and 4 in Sloan i'.  Photometric calibration of the data was obtained using measurements of field stars, calibrated as part of the Deep Lensing Survey (DLS; Wittman et al.~2002, Smith et al.~2002). 

Observations on UT 2010 June 6, 15, 18 and July 3 were obtained using the 0.9 m telescope of West Mountain Observatory, operated by Brigham Young University, in Utah.  We used the Finger Lakes PL-09000 charge-coupled device (CCD), which gives a 25.2\arcmin~field of view with 0.49\arcsec~pixels.  Images through a broadband Johnson-Cousins R filter were calibrated with reference to Landolt (1992) standard stars.

We also used the Keck 10 m diameter telescope atop Mauna Kea (altitude 4200 m) to observe P/Vales on UT 2010 August 20.  The Low Resolution Imaging Spectrometer (LRIS: Oke et al.~1995)  possesses independent blue and red channels separated by a dichroic filter.  We used the ``460'' dichroic (50\% peak transmission at 4900\AA~wavelength), and a broadband B filter on the blue side.  The B filter has central wavelength $\lambda_C$ = 4370\AA~ and Full Width at Half Maximum (FWHM) = 878\AA.  On the red side, we used a broadband R ($\lambda_C$ = 6417\AA, FWHM = 1185\AA) filter.   Photometric calibration of the data was secured using observations of  standard stars selected to have sun-like colors from the list by Landolt (1992) and cross-checked using field stars.  Unfortunately, the seeing was unusually poor and variable, in the range $\sim$2\arcsec~to 3\arcsec~FWHM.   As a result, we present only large aperture photometry from Keck, in order  to assess the total magnitude of P/Vales.  

\begin{figure}
\epsscale{1.0}
\plotone{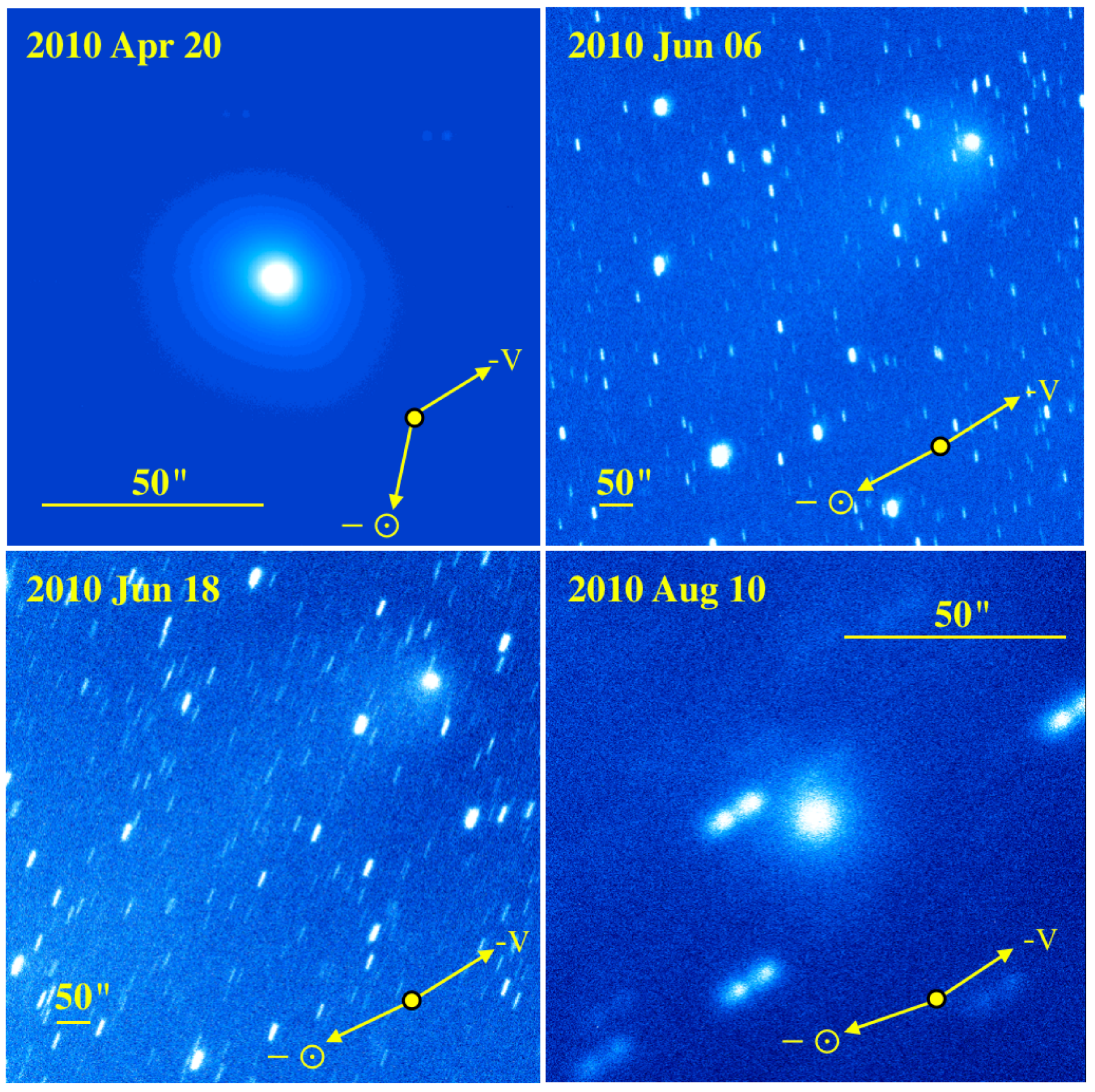}
\caption{Images of P/Vales on four dates.  Each panel has North to the top, East to the left, and is shown with a 50\arcsec~scale bar.  The projected anti-solar direction ($-\odot$) and the negative heliocentric velocity ($-V$) are shown as yellow arrows.  The image from UT 2010 June 15 is indistingusihable from that on June 18, and so not shown.  See Table (\ref{geometry}) for additional observational details.   
\label{images}}
\end{figure}

A timeline of observations is given in Table (\ref{geometry}) while representative composite images are shown in Figure (\ref{images}).

\subsection{Photometry}
We elected to measure photometry within a nested set of circular apertures having projected, fixed radii of 5$\times10^3$ km, 1$\times10^4$ km, 2$\times10^4$ km and 4$\times10^4$ km at the distance of P/Vales (Table \ref{photometry}).  The use of fixed linear (as opposed to angular) apertures ensures that we  measure the same volume around the nucleus independent of distance from the Earth, and obviates the need for a correction dependent on the surface brightness profile.  Sky subtraction was determined from the median signal measured in a contiguous annulus with inner radius 4$\times10^4$ km and extending out to (1 to 2) $\times 10^5$ km, depending on the dataset.    In some cases, we digitally removed background objects projected within the nested apertures.   We obtained photometric calibration of the data using   images of nearby Landolt (1992) standard stars and of field stars calibrated in the Pan STARRS survey (Tonry et al.~2012, Magnier et al.~2013).   Uncertainties on the photometry have several components.  The values listed in Table (\ref{photometry}) are the uncertainties determined empirically from the standard error on the mean of repeated measurements in each filter.  Small, additional uncertainties arise from the differences in the bandpasses  of the filters used at each telescope and, particularly for the Landolt stars, intrinsic uncertainties of the magitudes at the $\pm$0.01 to 0.02 magnitude level.  These  are all small compared to the uncertainty introduced by the unmeasured phase angle dependent darkening of P/Vales, and therefore of no consequence here.  All photometry was converted to standard Kron-Cousins BVR magnitudes using transformation equations from Smith et al.~(2002) for SDSS data and Tonry et al.~(2012) where Pan STARRS data were used.  
 
We focus our analysis on the  apparent R-filter magnitudes, $m_{R}$, which we convert into absolute magnitudes, $m_R(1,1,0)$, using

\begin{equation}
m_{R}(1,1,0) = m_{R} - 5\log_{10}(r_H \Delta) - \beta_{ph} \alpha.
\label{H}
\end{equation}

\noindent Here, $r_H$ and $\Delta$ are the instantaneous heliocentric and geocentric distances expressed in AU, respectively, and $\alpha$ is the phase angle in degrees.  Quantity $\beta_{ph}$ is the phase coefficient, equal to the ratio of flux densities scattered at angle $\alpha$ to $\alpha$ = 0\degr.  The phase coefficient is unmeasured in P/Vales but studies of other comets show that in back-scattering (small $\alpha$) geometries,  $\beta_{ph}$ = 0.02 magnitudes degree$^{-1}$ provides a useful approximation (Meech and Jewitt 1987).

The absolute magnitude provides a measure of the scattering cross-section, $C$ (km$^2$), through

\begin{equation}
C = \frac{1.5\times10^6}{p} 10^{-0.4 m_{R}(1,1,0)}
\label{C}
\end{equation}

\noindent in which $p$ is the geometric albedo.  We assume $p$ = 0.1 throughout, consistent with the nominal albedo of cometary dust (Zubko et al.~2017) and with the range of albedos (0.03 to 0.12)  inferred in the possibly similar, outbursting comet 17P/Holmes (Ishiguro et al.~2010).  The surfaces of Hildas have lower average albedos, $p \sim$ 0.05 (Grav et al.~2012). As we note later, ice was reported in P/Vales (Yang and Sarid 2010), suggesting that a higher albedo might be appropriate.  On the other hand,  ice becomes optically absorbing and dark unless very pure.  In short,  it is not obvious what the albedo of the ejected material should be, or even that albedo should be constant with respect to time since ejection.  Cross-sections can be easily scaled to other albedos from Equation (\ref{C}) by the factor $0.1/p$.


\subsection{Color Photometry}

\begin{figure}
\epsscale{0.9}
\plotone{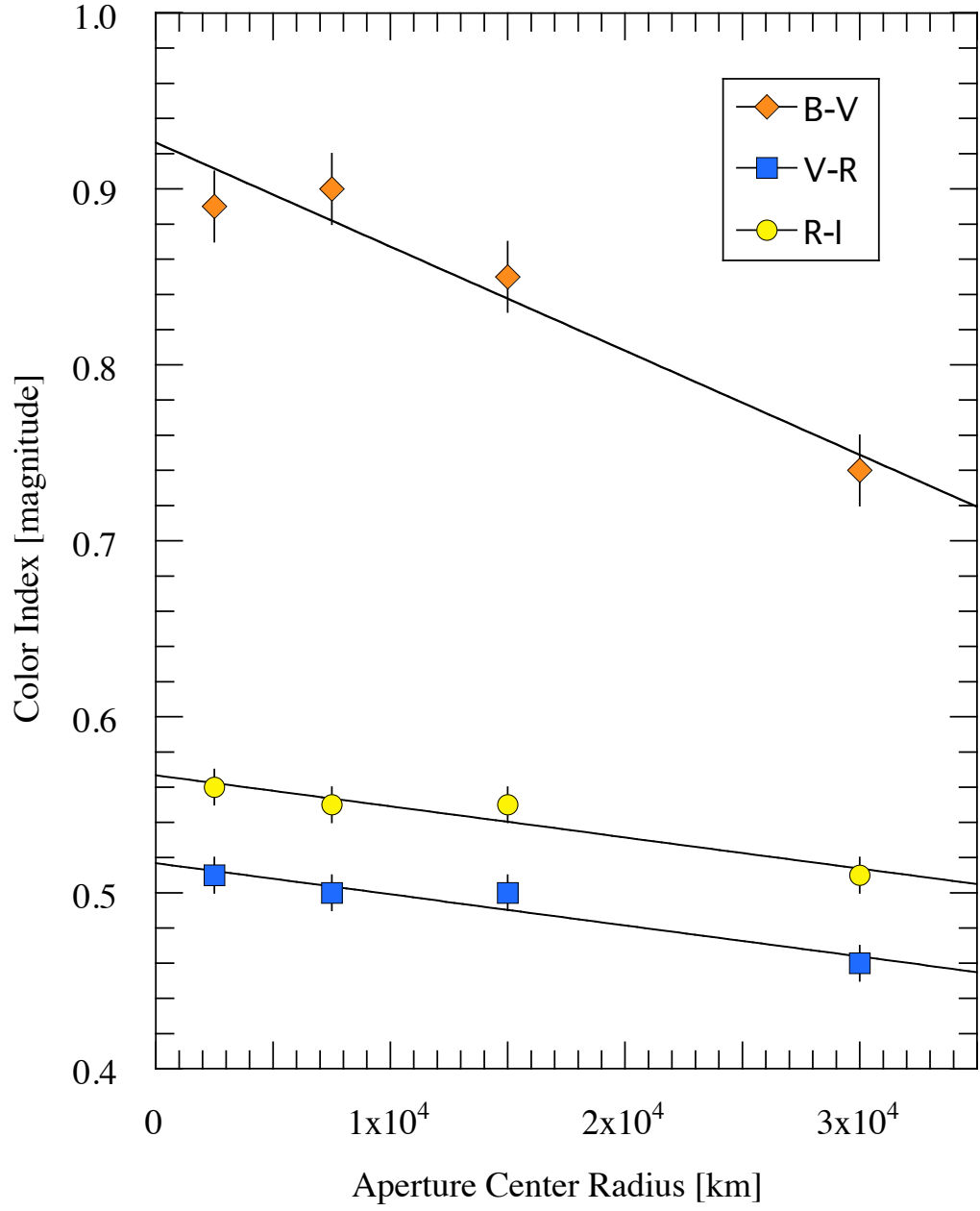}
\caption{Color indices as a function of projected aperture radius from UT 2010 April 20.
\label{colorplot}}
\end{figure}

Color measurements  from UT 2010 April 20 are presented in Table (\ref{colors}), along with data for the colors of the Sun from Holmberg et al.~(2006).  All three independent color indices, B-V, V-R and R-I, show evidence for a trend towards smaller values at larger radii, indicating that the outer parts of the coma are more blue than the inner parts (Figure \ref{colorplot}).  It is unlikely that the gradients are caused by gas contamination of the signal, both because  the resonance fluorescence bands of common molecules are weak and rarely detected at 3 AU and because these bands are largely confined to  wavelengths $\lesssim$5000\AA~and would mainly affect B-V.  All three color gradients gradients are individually significant at the 4$\sigma$ level, and may be related to particle fragmentation inferred from the surface brightness profile (Section \ref{profile}).  However, we do not model this effect here, since color depends on many unknown properties of the dust (size, composition, microstructure), any or all of which may vary with time in a transient body like P/Vales, and unique models of optical colors cannot be constructed.  Instead, we use the colors for comparative purposes only.  

The central colors are better representative of the source object and less likely to be affected by time-of-flight dependent optical effects in small particles.   Optically, the central (5000 km) colors of P/Vales are redder than the mean color of active short period comets (B-R = 1.40$\pm$0.02 vs.~1.22$\pm$0.02 for the comets; Jewitt 2015), but individual comets in the latter sample are scattered over the range B-R = 1.0 to 1.4, and some are as red as P/Vales.  The colors of P/Vales are much redder than the colors of  inactive Hildas. For example, the spectral slope across the B to R region is about $S'$ = 20\%/1000\AA, whereas the slopes measured for Hildas are all $S' \le$ 14\%/1000\AA~(Dahlgren and Lagerkvist 1995, Gil-Hutton and Brunini 2008). The red colors are more consistent with the mean colors of the nuclei of Jupiter family comets (for which B-V = 0.87$\pm$0.05, V-R = 0.50$\pm$0.03 and R-I = 0.46$\pm$0.03; Jewitt 2015).  This is not proof that P/Vales is a resonantly captured comet, but the colors are consistent with this interpretation.   A single large aperture (16.3\arcsec~radius, or 40,000 km) color measurement was obtained at Keck on UT 2010 August 10 under conditions of poor seeing, giving B-R = 1.29$\pm$0.10, consistent with the data from Table (\ref{colors}).

\subsection{Surface Brightness Profile}
\label{profile}
\begin{figure}
\epsscale{0.90}
\plotone{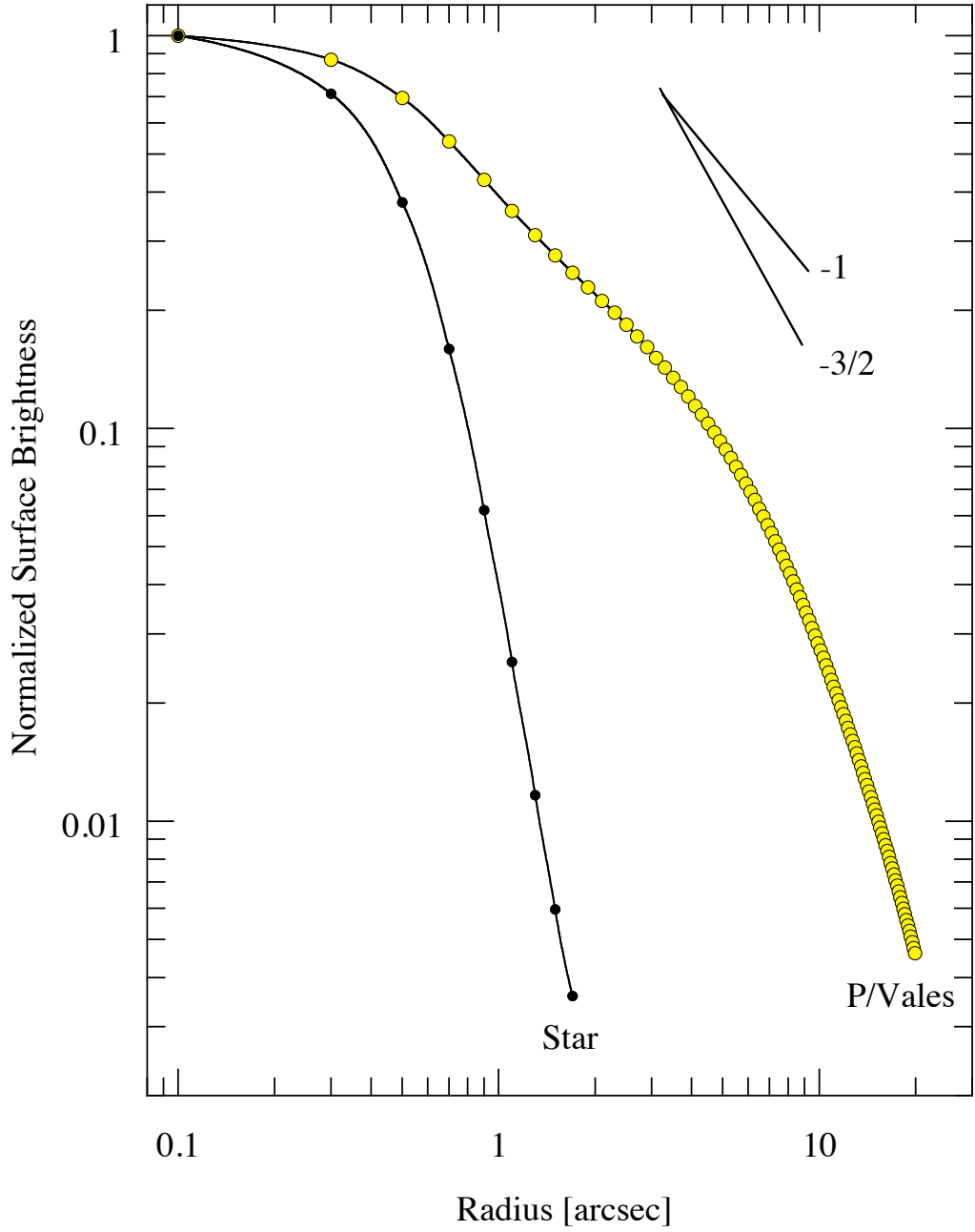}
\caption{Surface brightness profile of P/Vales on UT 2010 April 20 (yellow-filled circles) and a field star (black-filled circles).  Interpolated lines have been added to guide the eye.  In the upper right, line segments indicate logarithmic gradients $m$ = -1 and -3/2, as marked.  
\label{sb}}
\end{figure}

We measured the surface brightness profile, $\Sigma(\theta)$, where $\theta$ is the angular distance from the nucleus, in the UT 2010 April 20 data as follows.  The profile  was computed using a nested set of concentric annuli each 1 pixel (0.2\arcsec) wide, centered on the nucleus and with sky subtraction from a surrounding annulus having inner and outer radii 400 (80\arcsec) and 800 pixels (160\arcsec), respectively.  The profile was extracted out to a radius of 100 pixels (20\arcsec).  Similar measurements were taken to determine the point-spread function (PSF) from the profiles of field stars, using nearly simultaneous integrations of 10 s duration in which non-sidereal trailing is negligible.  The profiles are shown in Figure (\ref{sb}).

As expected, the coma profile shows a central excess caused by the convolution of the comet profile with the PSF.  At radius $\theta$ = 1\arcsec, however, the coma surface brightness is already $>10$ times larger than the PSF brightness and, at larger radii, the effects of the convolution can be ignored.   The profile becomes steeper at  angular radii $\theta \gtrsim$ 4\arcsec, as the physical edge of the coma is approached.  The logarithmic gradient of the surface brightness is $m = d\ln\Sigma(\theta)/d\theta$.  In the range 1\arcsec~$\le \theta \le$ 4\arcsec~(1545 to 6180 km at the comet), we find $m = -0.83\pm$0.01.  The gradient is significantly less steep than the canonical $m$ = -1 expected of a steady-state coma from the equation of continuity (Jewitt and Meech 1987).    This could be because the coma is itself not in steady state, with production at the nucleus varying on a timescale comparable to or shorter than the residence time for dust particles.  Alternatively,  the shallow gradient could reflect fragmentation of ejected particles, resulting in the creation of extra scattering cross-section (and perhaps a change in broadband color) as distance from the nucleus increases.  This was the  case in 17P/Holmes, where an even flatter surface brightness profile, $m$ = -0.27, was recorded (Stevenson and Jewitt 2012).  The fractional increase in the cross-section varies as $\theta^{1+m} - 1$.  With $\theta$ = 4\arcsec~and $m$ = -0.83, for example, the increase in the cross-section required to fit the gradient is  by a modest  $\sim$27\%.

The spectroscopic detection of water ice in the comae of both 17P/Holmes (Yang et al.~2009) and P/Vales (Yang and  Sarid 2010)  suggests a role for sublimation.  Specifically,  composite grains bound together by water ice would spontaneously disaggregate  through sublimation following their sudden expulsion from the nucleus into sunlight.  The ice sublimation rate is a strong function of the grain temperature as set by the heliocentric distance and albedo.  We solved the energy balance equation for an exposed, sublimating ice surface (c.f.~section 4.1 of Jewitt et al.~2020).  As an example, at $r_H$ = 3.11 AU, we calculate that the sublimation rate from an isothermal water ice sphere of albedo 0.1 is $f_s$ = 2$\times10^{-8}$ kg m$^{-2}$ s$^{-1}$.  The resulting sublimation lifetime of a grain of radius $a$ is $t_s \sim a \rho/f_s$.  By substitution, we estimate $t_s \sim$ 0.6 to 6 day, for $a$ = 1 to 10 $\mu$m.  Since $t_s$ is  comparable to the $\sim$4 day interval between the start of the outburst and the April 20 observation, we consider sublimation-induced disaggregation to be a plausible explanation of the shallow surface brightness gradient.  Spatially and/or temporally resolved spectroscopic observations of ice absorption could test this explanation.

\section{DISCUSSION}
\subsection{Analytic Considerations} 
\label{simple}

\textit{Nucleus Radius, $r_n$:} The strongest observational constraint on the nucleus radius is set by the non-detection at $m_V \ge$ 20 from the Catalina Sky Survey on UT 2010 April 15.4, when $r_H$ = 3.112, $\Delta$ = 2.131, $\alpha$ = 4.7\degr.   By Equation (\ref{H}) these values give absolute magnitude $m_V(1,1,0) \ge$ 15.8 and, by Equation (\ref{C}), a nucleus cross-section $C_n \le$ 7.2 km$^2$ and effective radius $r_n = (C_n/\pi)^{1/2} \le$ 1.5 km.  

\begin{figure}
\epsscale{0.8}
\plotone{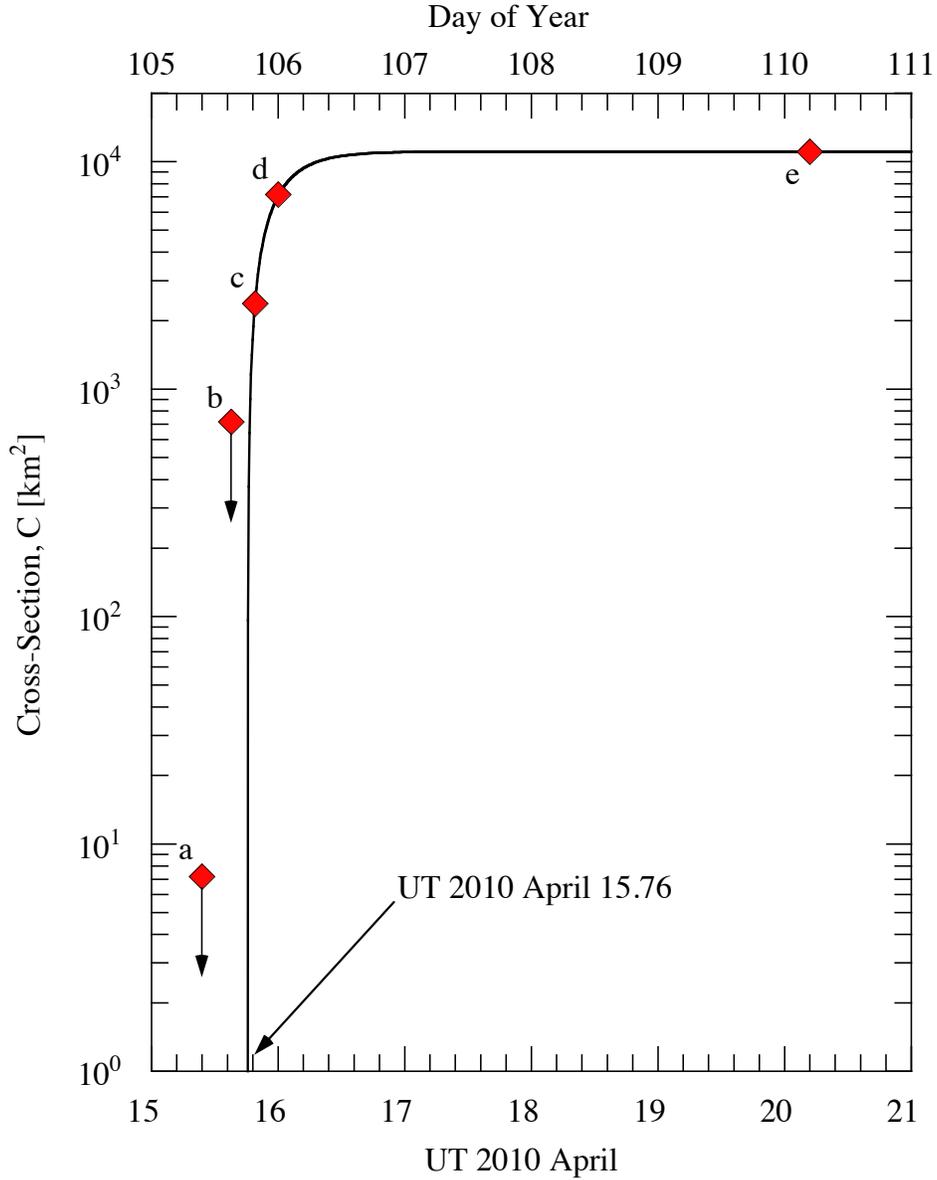}
\caption{Early-time lightcurve showing non-detections (down-pointing arrows)  from (a) Kowalski (reported in Vales et al.~2010) and (b) Balanutsa et al.~(2010) at UT 2010 April 15.4 and 15.56, respectively.  The first detection is (c) from April 15.82 (Balanutsa et al.~2010) and discovery is (d) at April 16.00 (Vales et al.~2010).    Point (e) shows our first observation from April 20.2 (Table \ref{geometry}).  The curve is an exponential fit to points (c), (d) and (e), indicating the start time UT 2010 April 15.76.  
\label{rise}}
\end{figure}

\textit{Start of Outburst, $T_0$:}  The non-detections on UT 2010 April 15.4 (Vales et al.~2010) and 15.56 (Balanutsa et al.~2010) and the first detection on April 15.82 strongly bracket the beginning of the outburst.  To obtain a better estimate of the start time within this range, we summarize the early-time observations in Figure (\ref{rise}), where we have converted the reported apparent magnitudes and limits to cross-sections, as described above.   Upper limits to the cross-sections are marked in the figure with down-pointing arrows.  We fitted the three detections with an exponential function $C = C_0(1- s \exp(-t/w))$, where $C_0$ = 11,069 km$^2$ is the peak cross-section determined on UT 2010 April 20.2 and $s$ and $w$  are constants.   The function, plotted in the figure, extrapolates to $C$ = 0 on UT 2010 April 15.76 (DOY 105.76) and provides our best estimate of the time of initiation.  This is only 0.2 days (5 hour) after the non-detection reported by Balanutsa et al.~(2010), 0.06 days (1.4 hour) before the first detection by the same observers and 0.25 days (6 hour) before the discovery.  Thus, even though the adoption of an exponential function is arbitrary, the data are highly constraining and there is little room for the start-time to be much different from that indicated by the fit.  The e-folding rise time of the cross-section given by the fit is $w^{-1}$ = 0.22 days (5.4 hour), corresponding to a rise-time half-life $t_{1/2} = \ln(2)w^{-1}$ = 3.7 hour.  Even shorter rise-times (0.3 to 0.6 hour) were reported in early-phase observations of 17P/Holmes (Hsieh et al.~2010).  The peak measured cross-section, $C_{max} = 1.8\times10^4$ km$^2$ on UT 2010 April 20.2 (Table \ref{photometry}), is equal to that of a circle of radius  $r \sim (C_{max}/\pi)^{1/2}$, or $r \sim$ 75 km.

\textit{Particle Parameters:} The detailed appearance of the coma is influenced by the time-profile of the emission and by the distribution of particle radii. We take the latter to be a differential power law with index, $q$, such that the number of particles with radius between $a$ and $a+da$ is $n(a) da = \Gamma a^{-q} da$ in the range $a_{min} \le a \le a_{max}$, with $\Gamma$ equal to a constant.  Before we apply a Monte-Carlo approach to constrain the dust parameters, it is informative to use basic physics to assess the nature of the particles.  We next obtain values for $a_{min}$, $a_{max}$ and $q$ to compare with values determined independently from a numerical model.

\textit{Lower Size Limit $a_{min}$:} We make two morphological observations from the images taken on UT 2010 April 20.2.   First, the coma extends in the sunward direction by a distance $\ell \sim 4\times10^7$ m.  We interpret $\ell$ as the turn-around distance for dust particles ejected sunward at speed $U$, and subject to a constant radiation pressure induced acceleration.  We write this acceleration as $\beta g_{\odot}$, where radiation pressure efficiency factor $\beta$ is a dimensionless number and $g_{\odot} = GM_{\odot}/r_H^2$ is the local gravitational acceleration to the Sun.   Parameter $\beta$ is a function of particle size, shape and composition.  As a useful approximation for dielectric spheres, we take $\beta \sim 10^{-6}/a$, where $a$ is the particle radius expressed in meters (Bohren and Huffman 1983).  Then, the equation of motion for a fixed acceleration gives

\begin{equation}
\beta  = \frac{2\ell r_H^2}{G M_{\odot} t^2}.
\label{beta}
\end{equation}

\noindent where  $G = 6.67\times10^{-11}$ N kg$^{-2}$ m$^2$ is the gravitational constant, $M_{\odot} = 2\times10^{30}$ kg  is the mass of the Sun and $t$ is the time of flight. 

Second, the comet possesses no clear radiation-swept tail (i.e.~dust particles launched sunward have been propelled to the anti-solar side of the nucleus by a distance no greater than $-\ell$).   Again, from the equation of motion for a fixed acceleration of $\beta g_{\odot}$, and substituting for $\beta$ from Equation (\ref{beta}),  we find

\begin{equation}
U = \frac{2\ell}{t} 
\label{U}
\end{equation}

%

\noindent For the Magellan observation we set $t$ = 4.45 days (3.8$\times10^5$ s) and $\ell = 4\times10^7$ m to find $U$ = 210 m s$^{-1}$ and $\beta$ = 0.9 (implying $a \sim$ 1.1 $\mu$m).  Strictly, $\ell$ is a lower limit to the true turn-around distance because of the effects of projection into the plane of the sky. Therefore, the derived values of $U$ and $\beta$ are also lower limits.  For present purposes, however, these crude estimates are sufficient to show that  the early-stage morphology of the envelope of the coma  is controlled by fast-moving particles with radii $a \sim$ 1 $\mu$m.    

\begin{figure}
\epsscale{1.0}
\plotone{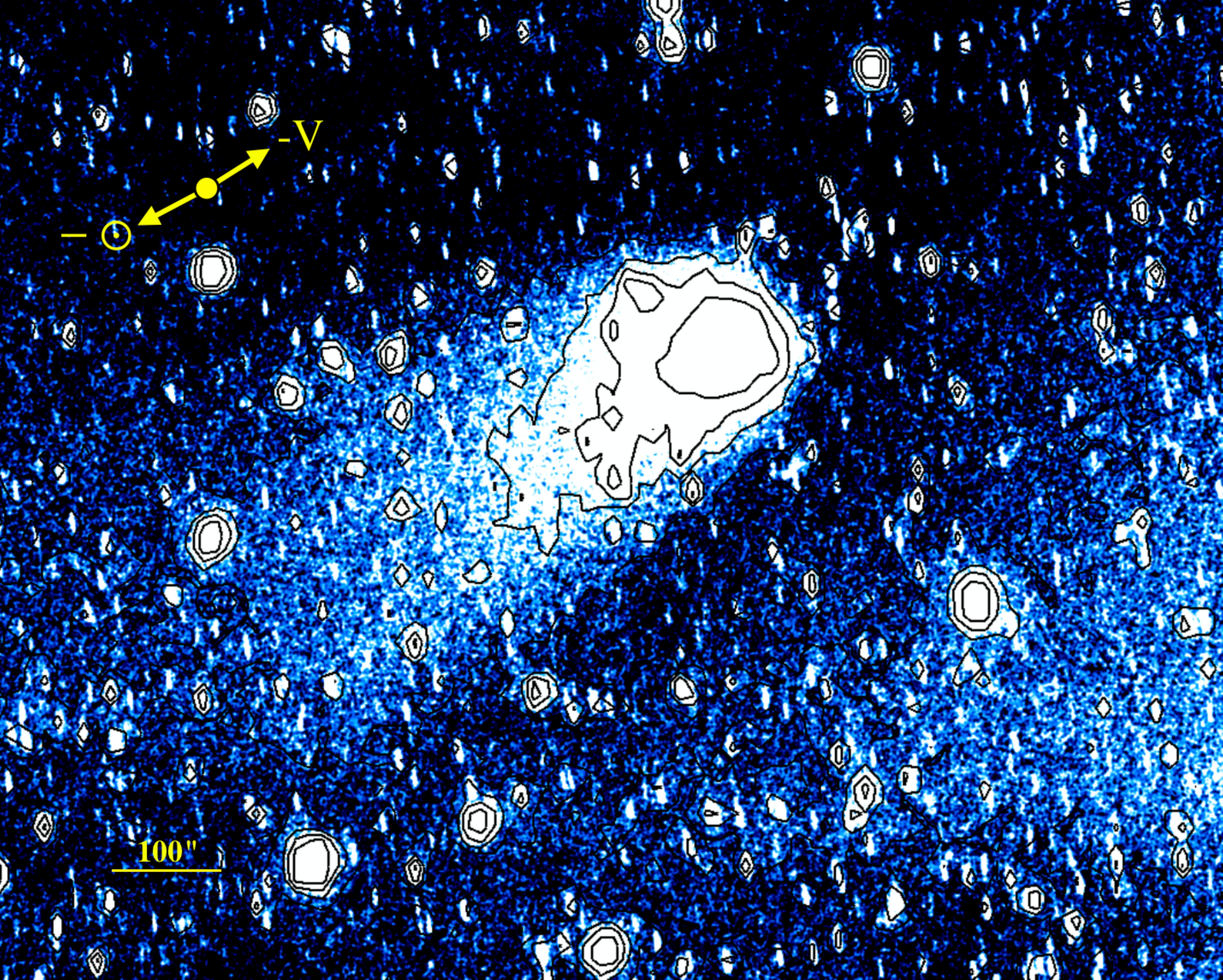}
\caption{Hard stretch of P/Vales on UT 2010 June 06, with North to the top, East to the left, and  a 100\arcsec~scale bar.  The projected anti-solar direction ($-\odot$) and the negative heliocentric velocity ($-V$) are shown as yellow arrows.  The image has been smoothed by convolution with a gaussian of 1\arcsec~FWHM to enhance faint structure.  Light in the lower right is internally scattered from a bright field star.  The tail extends to the edge of the image.  
\label{jun06}}
\end{figure}

By UT 2010 June 6 ($t$ = 52 days after the outburst), the tail of P/Vales can be traced to the edge of the field of view, a distance $\ell \ge 1.3\times10^9$ m (Figure \ref{jun06}).  Application of Equation (\ref{beta}) then gives $\beta \ge$ 0.3 for the dust particles at the edge of the field of view.  Again, the large-scale morphology on this date is controlled by  the dynamics of micron-sized particles accelerated by  solar radiation pressure.  Large particles, less easily accelerated by radiation pressure,  must be located closer to the nucleus on this date.  The absence of a distinct feature attributable to large particles (specifically, a dust trail in data from UT 2010 June 6 (Figure \ref{images}), when the Earth was only 1.1\degr~above the orbital plane of P/Vales (Table \ref{geometry})), shows that the large particle contribution to the cross-section of the ejected material is minor on this date.

\begin{figure}
\epsscale{0.85}
\plotone{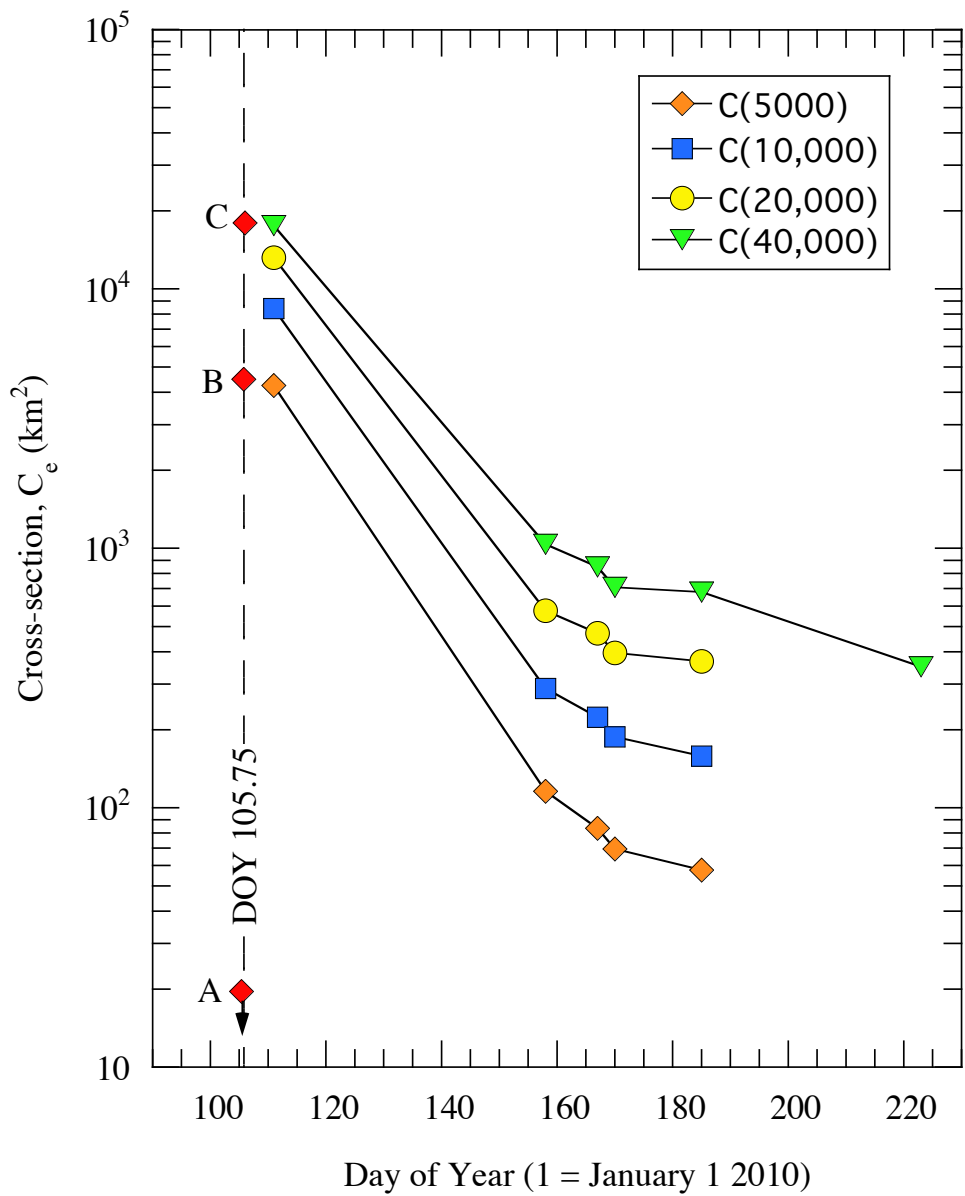}
\caption{Scattering cross-section as a function of Day of Year in 2010.  Red circles show A: an upper limit to the cross-section set by pre-discovery data (reported in Vales et al.~2010), B: a pre-discovery detection in the rise phase from Balanutsa et al.~(2010) and C: the discovery (Vales et al.~2010). Other data from Table (\ref{photometry})  Black lines connect data from each aperture.  The date of outburst initiation is shown as a vertical dashed line.
\label{Ce_vs_DOY}}
\end{figure}

\textit{Upper Size Limit $a_{max}$:} Larger particles are less well-coupled to the outrushing gas and will have smaller terminal speeds.  In gas drag expulsion, the terminal speed of a dust grain is related to its radius by $U(a) = U_1(a_0/a)^{1/2}$, where constant $U_1$ is the speed of a particle at reference radius $a_0$.  We take $a_0$ = 10$^{-6}$ m and $U_1$ = 210 m s$^{-1}$.  To estimate the radius of the largest (slowest) particle that can be ejected against the gravity of the nucleus (ignoring possible effects of cohesion), we set  $U = V_e$, where $V_e$ is the gravitational escape speed, to find   

\begin{equation}
a_{max} = \frac{3 a_0 U_1^2}{8 \pi G \rho r_n^2}
\label{amax}
\end{equation}

\noindent where  we have assumed for simplicity that the nucleus is spherical and of density $\rho$. Substituting $r_n $ = 1.5 km, we find from Equation (\ref{amax})  the largest ejectable dust radius,  $a_{max} \sim$ 6 cm.  These large, low speed particles would take a time $t \sim  3\times10^7$ s (1 year) to cross the 40,000 km photometry aperture, consistent with the persistence of excess cross-section in the lightcurve (Figure \ref{Ce_vs_DOY}) on 100 day timescales.


\textit{Size Distribution Index, $q$:} Smaller, faster dust particles escape the photometry aperture more quickly than larger, slower ones.  Therefore, in an impulsive outburst, the mean size of the particles in the aperture should increase with time because of the preferential loss of small particles, even as the total cross-section decreases.  An expression for the fading caused by the selective loss of particles from a photometry aperture was derived by  Jewitt et al.~(2017).  Here, we modify this expression to consider particles within an annulus, rather than a circular aperture, in order to compare with annular data extracted from Table (\ref{photometry}).

Consider a photometry annulus with inner and outer radii, $r_1$ and $r_2$, respectively, and a measurement taken time $t$ after the ejection.  Particles traveling more slowly than $U_{min} = r_1/t$ will not have reached the inner edge of the annulus, while those traveling faster than $U_{max} = r_2/t$ will have escaped its outer edge.   Substituting for the velocity-size relation, $U(a) = U_1(a_0/a)^{1/2}$,  we find the critical radii 

\begin{equation}
a_1(t) = a_0 \left(\frac{U_1 t}{r_1}\right)^2~~\textrm{and}~~a_2(t) = a_0 \left(\frac{U_1 t}{r_2}\right)^2.
\end{equation}

\noindent Particles with radii $a < a_2(t)$ will have escaped the outer edge of the annulus after time of flight, $t$, while those with $a > a_1(t)$ travel so slowly that they have not yet reached the inner edge.  Therefore, at any time, $t$, the particles within the annulus are  confined to the radius range $a_2(t) < a < a_1(t)$, assuming impulsive emission

The fraction of the ejected dust cross-section remaining within the annulus at time $t$ since ejection is then

\begin{equation}
\frac{\Delta C(t)}{C(0)} = \frac{\int_{a_2}^{a_1}\pi a^2 n(a) da}{\int_{a_{min}}^{a_{max}}\pi a^2 n(a) da}
\end{equation}

\noindent where $C(0)$ is the total ejected cross-section.  Substituting $n(a)da = \Gamma a^{-q}da$ and evaluating, we obtain

\begin{equation}
\frac{\Delta C(t)}{C(0)}  = \left(\frac{a_0}{a_{min}}\right)^{3-q}  \left[\left(\frac{1}{r_1}\right)^{6-2q} - \left(\frac{1}{r_2}\right)^{6-2q}\right]   (U_1 t)^{6-2q}     
\label{model}
\end{equation}

\noindent where we have assumed $a_{max} \gg a_{min}$, $q > 3$ and that the size range $a_1 - a_2$ is contained with $a_{min} - a_{max}$.  The  time dependence in Equation (\ref{model}), $\Delta C(t)/C(0) \propto t^{6-2q}$, is the same as derived in Jewitt et al.~(2017), and allows us to estimate $q$ from the time-dependent annular photometry.

\begin{figure}
\epsscale{0.85}
\plotone{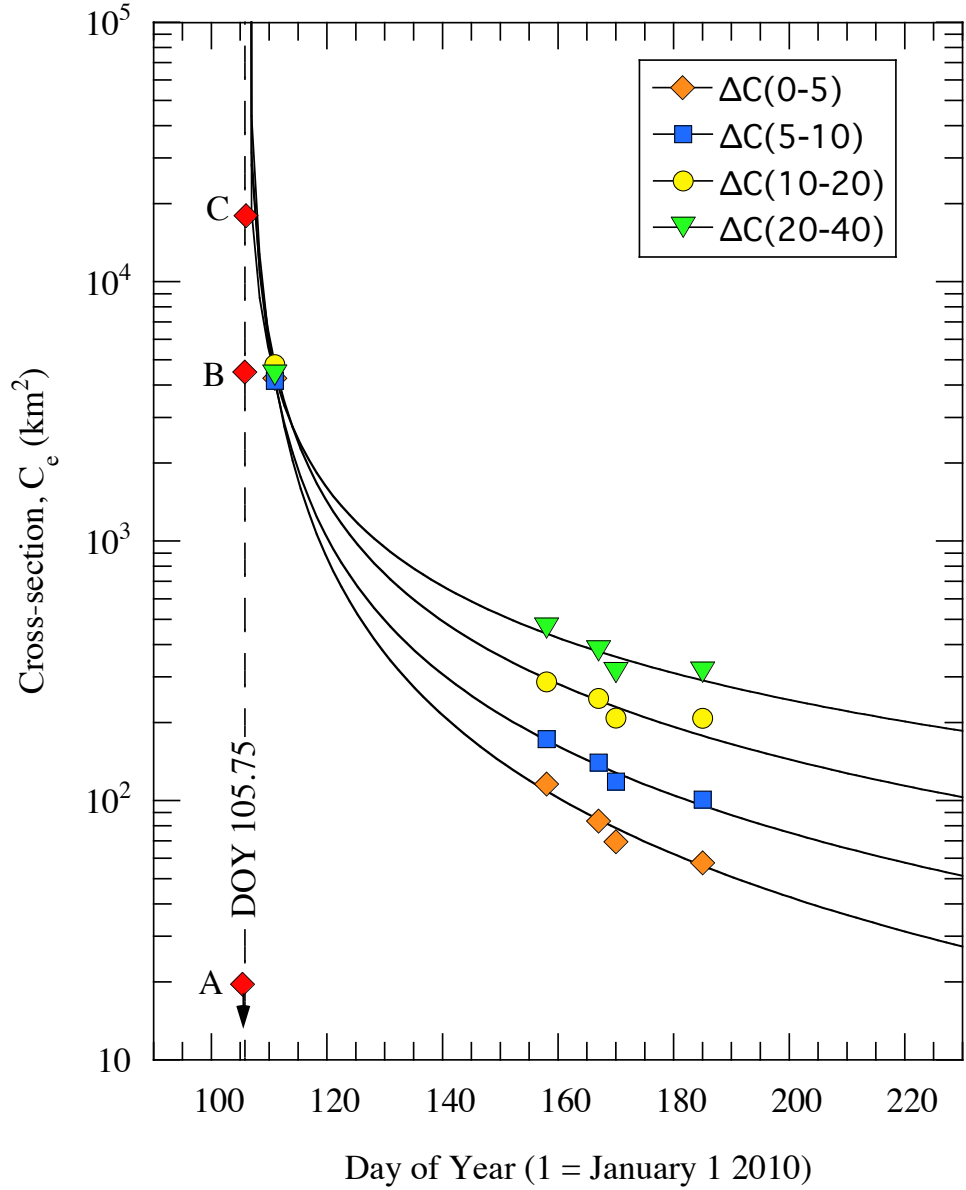}
\caption{Annulus scattering cross-section vs.~Day of Year.  The black lines show least-squares fits to the data, yielding a particle size distribution index (differential) of $q$ = 3.58, 3.64, 3.75 and 3.70, from top to bottom.  Other features are the same as in Figure (\ref{Ce_vs_DOY}).
\label{comafit}}
\end{figure}

The four annular cross-sections, $\Delta C(t)$, are plotted in Figure (\ref{comafit}).  We least-squares fitted power laws to $\Delta C(t)$  in order to find the index, $q$ from Equation (\ref{model}).  Lines in the figure show independent fits to photometry from the four apertures presented in Table (\ref{photometry}).  For the $\Delta C$ = 0-5,000 km, 5,000-10,000 km, 10,000-20,000 km, 20,000-40,000 km apertures we find, respectively, $q$ = 3.80, 3.70, 3.61 and 3.50.  The formal uncertainties on these fits are $\pm$0.01 to $\pm$0.02, but we use the standard error on the mean of the four values as a better measure of the true scatter.  The mean size distribution index from fits to all four apertures is $q$ = 3.61$\pm$0.06.  Small differences between the values could result from many causes (perhaps the size distribution is not a power law, as assumed in Equation (\ref{model}), perhaps the emission was not impulsive or the size-speed relation is invalid, perhaps the grain albedo changes with time as ice sublimates away).  Given these many potential problems, the overall  agreement between the four independent determinations of $q$ is good.  The derived value is close to $q$ = 3.7$\pm$0.1 determined for fragmenting asteroid P/2012 F5 (Gibbs) (Moreno et al.~2012) and $q$ = 3.6$\pm$0.6 for fragmenting comet 332P/Ikeya-Murakami (Jewitt et al.~2016). 


\textit{Dust Mass, $M$:} The mass of an optically thin collection of spheres is related to the sum of their cross-sections by 

\begin{equation}
M = \frac{4}{3}\rho \overline{a} C
\label{mass}
\end{equation}

\noindent  where $\overline{a}$ is the area-weighted mean particle size responsible for cross-section, $C$.  The latter is given by 

\begin{equation}
\overline{a} = \frac{\int_{a_{min}}^{a_{max}}a^3 \Gamma a^{-q} da}{\int_{a_{min}}^{a_{max}}a^2 \Gamma a^{-q} da}
\end{equation}

\noindent which, for $q$ = 3.61 simplifies to

\begin{equation}
\overline{a} = 1.56 a_{max}^{0.39} a_{min}^{0.61}
\label{abar}
\end{equation}.  

\noindent Substitution gives $\overline{a} = 1\times10^{-4}$ m (100 $\mu$m).  

We see from  Table (\ref{photometry}) and Equations (\ref{mass}) and (\ref{abar})  that the peak dust cross-section, $C$ = 17.6$\times10^3$ km$^2$, with $\rho$ = 500 kg m$^{-3}$, gives dust mass $M = 1.2\times10^{9}$ kg.   This is about 10$^{-4}$ times the mass of the nucleus, considered as a sphere of the same density and radius 1.5 km.  The bulk of this mass was released over a period $\tau \le$ 1 day (Figure \ref{rise}), corresponding to a  mean mass production rate $M/\tau \gtrsim$ 13,600 kg s$^{-1}$.   This rate is comparable to the (1 to 2)$\times10^4$ kg s$^{-1}$ measured in super-active comet C/1995 O1 (Hale-Bopp) at similar heliocentric distance (Weiler et al.~2003), although it is sustained in P/Vales for only a short period of time.   

\textit{Specific Energy, E/M:}
The energy per unit mass of ejecta is $E/M = (1/2) U^2(\overline{a})$.  For a $q$ = 3.6 distribution, the energy is carried by the small (fast) particles, because of the $U^2$ dependence, while the mass is carried by the large particles.  With $\overline{a}$ = 100 $\mu$m, $U(\overline{a}$) = 21 m s$^{-1}$ and $E/M$ = 220 J kg$^{-1}$.  This specific energy is about 10$^2$ times smaller than the corresponding quantity estimated for comets 332P/Ikeya-Murakami and 17P/Holmes.  The total energy of the outburst is $E \sim 3\times10^{11}$ J, an order of magnitude smaller than the outburst of 332P/Ikeya-Murakami and at least 10$^3$ times smaller than that of 17P/Holmes (summarized by Ishiguro et al.~2016).

\subsection{Monte-Carlo Model}
\label{MC}
We applied a 3D Monte-Carlo simulation
(Ishiguro et al.~2007) to P/Vales. In the model the motions of dust particles,  under the action of solar gravity and radiation pressure, are followed as a function of particle size and time and direction of ejection.   Our aim is not to reproduce the comet exactly since, given the number of free parameters in the model, a match can almost always be obtained.   Instead, we aim to provide a consistency check of the results deduced analytically in Section (\ref{simple}).  

As in Section (\ref{simple}), we set the dust velocity-size relation, $U(a) = U_1(a_0/a)^{1/2}$, as expected of gas drag acceleration, with $U_1$ = 210 m s$^{-1}$.  We further assumed the range of particle sizes 10$^{-6} \le a \le 6\times10^{-2}$ m, a differential power-law index $q$ = 3.6 and impulsive ejection on UT 2010 April 15.76.    The model includes, as a free parameter, the angular dependence of the dust production rate, represented by  a cone having its apex at the center of the nucleus, the Sun on its axis and a variable half-width, $w$. By experimentation, we found that   $w$ is most strongly constrained by the angular width of the dust tail in data from 2010 June, and that there is a trade-off between $w$ and $U_1$ in the dust speed vs.~size relation.  Given the nominal speed $U_1$ = 210 m s$^{-1}$, we found that $w$ = 25\degr~provided an acceptable fit to the tail.  Values $w$ = 15\degr~and $w$ = 35\degr~produced tails, respectively, too narrow and too wide to fit the images from 2010 June.  

\begin{figure}
\epsscale{1.00}
\plotone{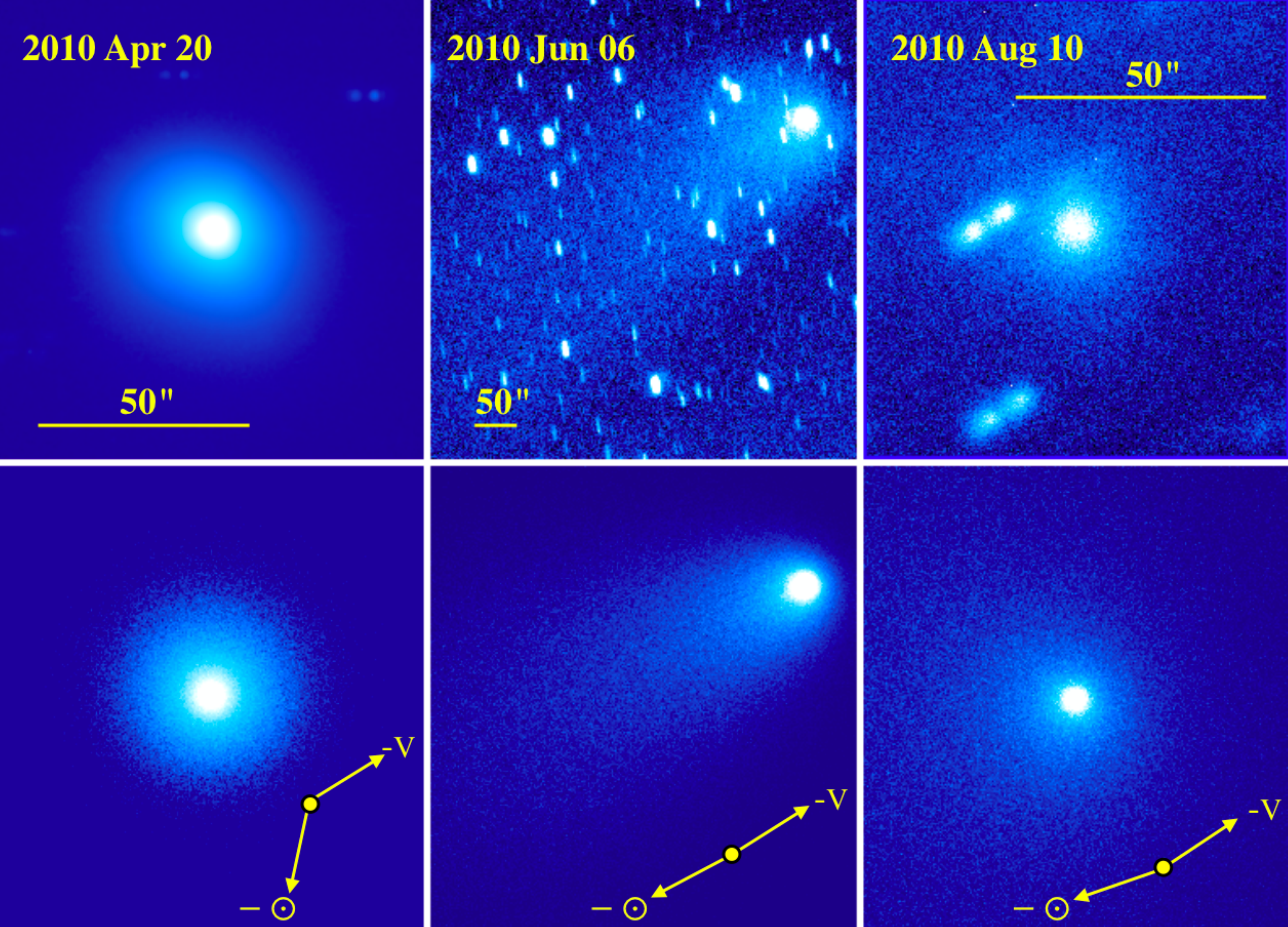}
\caption{Monte Carlo simulations (bottom row) of P/Vales compared with images (top row) on three dates.  Each panel has North to the top, East to the left and scale bars and direction arrows as in Figure (\ref{images}).
\label{mc}}
\end{figure}

Examples of the Monte Carlo simulations are shown in Figure (\ref{mc}), where it is seen that the parameters deduced analytically indeed generate models that match the data well. 
For example, the  simulations  show that, with the parameters deduced in Section (\ref{simple}), no large-particle trail can be discerned even near the crossing of the orbit plane on UT 2010 June 6 (Table \ref{geometry}).  This is consistent with the absence of a trail in the data (Figure \ref{mc}), confirming that large, slowly ejected grains present only a small fraction of the total cross-section even when spatially dispersed in the plane of the sky by radiation pressure. The simulations also reproduce the persistence of near-nucleus dust into 2010 August and the roughly circular late-stage appearance of the comet (Figure \ref{mc}).  This confirms that, by August 10,  only the largest, slowest particles remain in the vicinity of the nucleus and the smaller particles, constituting the elongated tail in data from 2010 June, have been swept away.  Overall, Monte Carlo simulations based on  the parameters deduced in Section \ref{simple} provide an acceptable match to the broad-brush appearance of P/Vales.

\begin{figure}
\epsscale{1.00}
\plotone{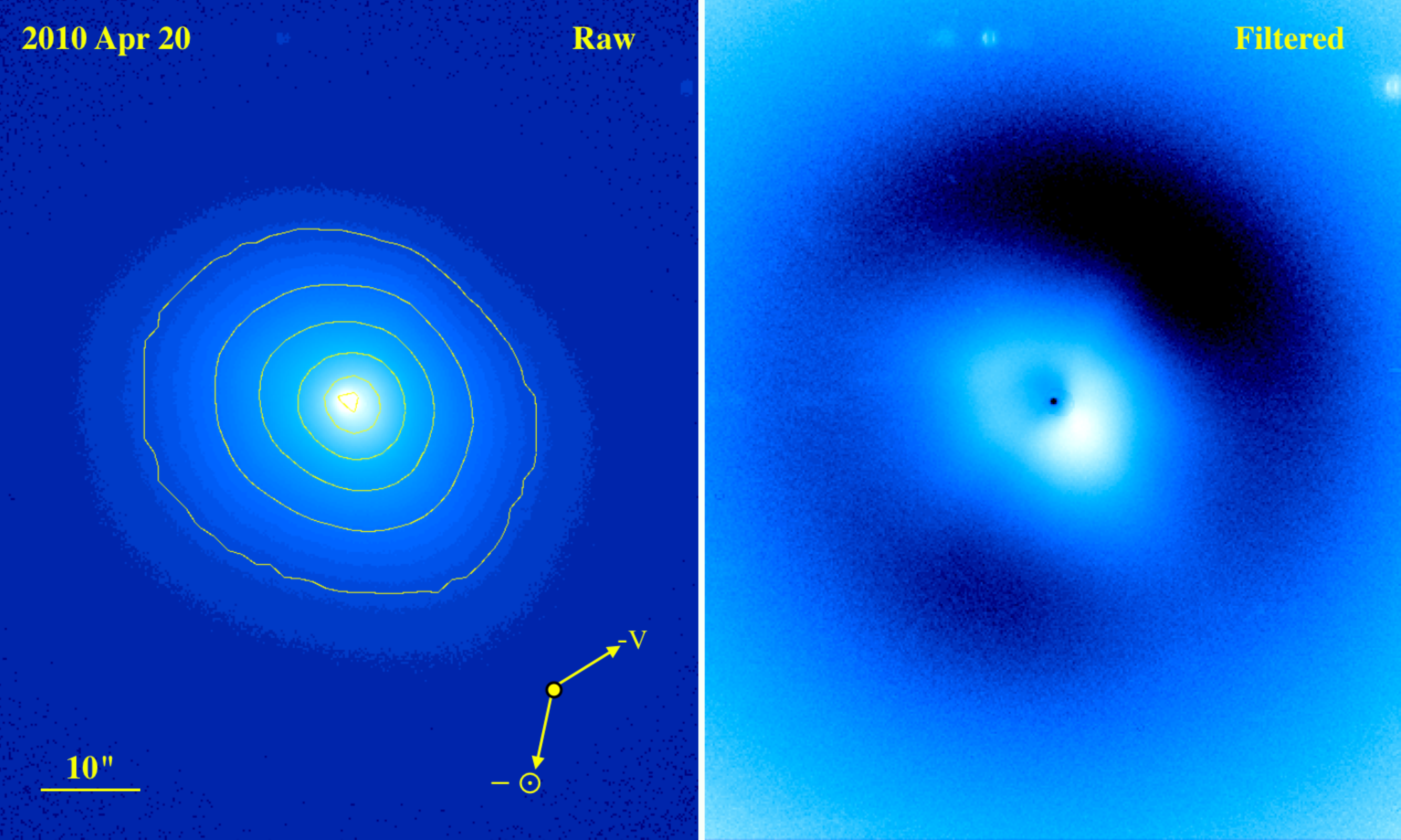}
\caption{(left) Image from UT 2010 April 20 (same as in upper left panel of Figure \ref{images}) and (right) this image divided by a model in which surface brightness varies inversely with distance from the nucleus, in order to suppress radial variations and enhance azimuthal ones.  The enhanced panel, in which the location of the nucleus is marked by a black dot, clearly shows an excess of coma material to the south west.  In the left panel, the outermost contour is at 2.5 arbitrary units and successive contours are each brighter by a factor of two. A 10\arcsec~scale bar and direction arrows are shown.  North is to the top, East to the left.
\label{oneoverrho}}
\end{figure}

Closer examination of the data shows evidence for unmodeled anisotropy in the coma in addition to that caused by solar radiation pressure.  In Figure (\ref{oneoverrho}) we show, on the left, the image from UT 2010 April 20 and, on the right, the same image divided by a normalized profile in which the surface brightness varies inversely with angular distance from the nucleus (the location of which is marked in the figure by a black dot; Samarasinha and Larson 2014).  The filtered image clearly shows excess emission near position angle 220\degr, and a broad deficit towards 330\degr.  The amplitudes of these features are small (explaining why they are not evident in the left, unfiltered, image) but they confirm that mass loss from P/Vales was not isotropic.  We have not attempted to model the anisotropy evident in  Figure (\ref{oneoverrho}).

\subsection{Outburst Mechanism}  We briefly consider possible causes of the outburst of P/Vales. 

The existence of high speed ejecta, with $U \ge 210$ m s$^{-1}$ for 1 $\mu$m particles,  strongly suggests the action of gas drag.  To see this, we note that, at $r_H$ = 3.112 AU, the local isothermal blackbody temperature is $T_{BB}$ = 158 K while the mean thermal speed of H$_2$O molecules at this temperature is $V_{th} = (8 k T_{BB}/(\pi \mu m_H))^{1/2}$, where $\mu$ = 18 is the molecular weight of water and $m_H = 1.67\times10^{-27}$ kg is the mass of the hydrogen atom.  We compute $V_{th}$ = 430 m s$^{-1}$.  The observation that, within a factor of two, $U \sim V_{th}$ for micron-sized dust grains is consistent with their acceleration by gas drag, as is the case in comets, generally.

\textbf{Volatile Explosion:}   
Evidence for the interplay between cometary activity and dynamics is well-established.  For example, the largest values of the non-gravitational accelerations of comets (which provide a measure of the mass loss rate) are correlated with recent,  inward migration of the perihelion distance (Rickman et al.~1991).   Could it be that the recent trapping of P/Vales into its quasi-Hilda orbit has triggered either enhanced sublimation of sub-surface volatiles or the exothermic crystallization of buried amorphous ice, leading to pressure build-up and eventual rupture of  a cohesive mantle (c.f.~Samarasinha 2001)?  

In this scenario, the  delay between the  Jupiter encounters of the previous century and the  outburst in 2010 would  be a consequence of the slow conduction of heat from the surface to the buried ice.   Solution of the conduction equation shows that the distance, $d$, over which heat is conducted in time $\tau$ is approximately $d \sim (\kappa \tau)^{1/2}$, where $\kappa$ is the thermal diffusivity of the material.  The diffusivity of solid dielectric materials is $\kappa \sim 10^{-6}$ m$^2$ s$^{-1}$, but porosity reduces $\kappa$ significantly. For example, laboratory measurements of highly porous dielectric powders give $\kappa \sim 10^{-9}$ to $10^{-8}$ m$^2$ s$^{-1}$ (Sakatani et al.~2018).  If we suppose that the most recent inward excursion of the perihelion distance occurred in 1973, the resulting $\tau =$ 35 years (10$^9$ s) delay would correspond to a buried ice depth $d \sim$ 1 m to 3 m.  Earlier perihelion decreases would correspond to larger $\tau$ and deeper ice, but the diffusive $d \propto \tau^{1/2}$ dependence suggests that  $d$ is unlikely to be substantially larger than $\sim$10 m.

The Monte-Carlo models (Section \ref{MC}) indicate ejection into a cone of half-angle $w \sim$ 25\degr, amounting to a solid angle $\Omega  \sim$ 0.2 steradian.   Given this, we estimate the area of the region producing the outburst as $A_{ice} = 4\pi r_n^2 (\Omega /4\pi)$  and the thickness of ice needed to provide the ejecta mass, $M$, as $d_{ice} \sim M/(\rho \Omega r_n^2)$.  Setting  $M = 1.2\times10^9$ kg, $\rho$ = 500 kg m$^{-3}$ and $r_n \le 1.5$ km, we find  $A_{ice} \le$ 0.4 km$^2$ (1.5\% of the nucleus surface) and $d_{ice} \ge$ 5 m.  Thus, a volatile pocket  modest in both thickness and areal extent  could supply the ejecta responsible for the outburst in P/Vales.    The crystallization of amorphous water ice releases $\sim10^5$ J kg$^{-1}$, far more than the average $\sim$220 J kg$^{-1}$ measured for the specific energy of the ejecta.   Thus, even a very inefficient conversion of crystallization energy into kinetic energy would be sufficient to drive the outburst and we consider a delayed explosion caused by slowly conducted heat to be a plausible explanation for the outburst of P/Vales.  On the other hand, whether or not amorphous ice could persist at meter depths in P/Vales depends on its past dynamical history which is, as with all comets,  chaotic and unknown.    We note that the spectrum reported by Yang and Sarid (2010) showed the 1.65 $\mu$m wavelength absorption indicative of  water ice in the crystalline state.  However, at the local blackbody temperature (158 K at 3.1 AU), crystallization occurs in $\ll$1 second, and the initial state of the ice before ejection into sunlight cannot be spectroscopically ascertained.

\textbf{Cliff Collapse:}   Could the outburst of P/Vales have resulted from the collapse of a cliff or an overhang on the nucleus surface?   In this scenario, the collapsed material crumbles and spreads across the surface as a landslide from which the entrained volatiles  sublimate, expelling dust and debris.  A high-temperature Terrestrial analog might be found in the pyroclastic flows produced by gas-rich, high viscosity volcanic magmas, where ejected solids degas and move as a ground-hugging, dense flow.  A cliff-collapse outburst of $\sim10^6$ kg  has been observed close-up on the nucleus of 67P/Churyumov-Gerasimenko (Pajola et al.~2017).      

While qualitatively attractive, cliff collapse offers a less obvious explanation of the much larger ($\sim10^9$ kg) outburst in P/Vales.  To see this, we first note that the area $A_{ice}$, estimated above, corresponds in this scenario to the area of the landslide needed to generate the outburst by sublimation.
We solved the energy balance equation for sublimating carbon monoxide (CO) ice, considered as a representative supervolatile, finding that at $r_H$ = 3.112 AU, the specific sublimation rate of a perfectly absorbing CO ice surface oriented normal to the Sun is $f_s(CO) = 4.6\times10^{-4}$ kg m$^{-2}$ s$^{-1}$.   The sublimation rate from the landslide is then just $dM/dt = f_s(CO) A_{ice}$.  By substitution, $dM/dt \sim 180$ kg s$^{-1}$.  This is an upper limit to the true $dM/dt$ because the ice is unlikely to be perfectly absorbing (Pajola et al.~2017 determined an albedo $\sim$0.4 in their example) and because the landslide surface  is unlikely to be oriented perpendicular to the direction of the Sun.  Significantly, $dM/dt$ from this estimate is two orders of magnitude smaller than the  actual production rate obtained from the rising lightcurve,  $M/\tau = 13,600$ kg s$^{-1}$ (Section \ref{simple}). Sublimation from a fresh landslide produced by cliff collapse cannot directly supply the massive outburst of P/Vales.

However, expulsion of ice-containing debris from the landslide surface into an expanding, avalanche-like cloud having cross-section $\gg A_{ice}$ could strongly amplify the production rate.  For example, we found that the peak measured cross-section is equal to that of a circle of radius  $r \sim (C_{max}/\pi)^{1/2}$, or $r \sim$75 km (Section \ref{simple}).     Ice in these particles would then sublimate at a rate $\sim \pi r^2 f_s(CO)$ which, with $r$  = 75 km, could supply $dM/dt \sim 8\times10^6$ kg s$^{-1}$, two orders of magnitude larger than required by the data.   Furthermore, if the cliff collapse were to expose amorphous ice, the resulting immediate exothermic crystallization could easily supply enough energy to drive the outburst.  Particles with the mean radius $\overline{a}$ = 100 $\mu$m have velocity $U \sim$ 20 m s$^{-1}$, and would take $\sim$1 hour to fill an optically thin hemisphere of radius 75 km, a timescale consistent with the short rise-time of the lightcurve (Figure \ref{rise}).  The coma was probably optically thin even at the time of the first detection (April 15.82, about 0.06 days after the best-fit initiation time on April 15.76, corresponding to a delay of $\sim$1.5 hours).  We conclude that cliff collapse of a sufficient volume could generate an outburst having the magnitude and the rapid rise-time implied by the observations.

An unresolved issue with the cliff collapse hypothesis is one of timing.  Why, other than by coincidence, would cliff collapse occur decades after the entrapment of P/Vales into the 3:2 mean-motion resonance?  Perhaps low-level activity driven by the sublimation of near-surface water ice in the years prior to outburst caused incipient instability leading to collapse.  In the absence of relevant observational evidence, however, we can say nothing about this possibility.

\textbf{Impact:}
High speed ejecta can also be produced by impact.  However, two factors make an impact origin unlikely.  First, the collision probability in the Hilda population ($P_i \sim 2\times10^{-18}$ km$^{-2}$ yr$^{-1}$; Dahlgren 1998, dell'Oro et al.~2001)  is lower than in the main-asteroid belt ($P_i \sim 5\times10^{-18}$ km$^{-2}$ yr$^{-1}$, Bottke and Greenberg 1993). The Hilda population is $\lesssim$1\% of  the main-belt population and the quasi-Hilda population is orders of magnitude smaller still.  Second, and more seriously, the quasi-Hildas have short dynamical lifetimes in the 3:2 resonance region (e.g.~10$^3$ - 10$^4$ year, Gil-Hutton \& Garc{\'\i}a-Migani 2016). The likelihood of a substantial  impact in this small population with a tiny residence window is very small.  In the particular case of P/Vales, impact within a few decades, or even centuries, of its  injection into the present orbit (Marsden 2010) is incredibly unlikely.  

\textbf{Other Processes: } The active asteroids are driven by several low energy processes in addition to sublimation gas-drag and impact (Jewitt 2012, Jewitt et al.~2015).  These include rotational disruption of the parent body, cracking caused by thermal expansion and/or desiccation stresses, and electrostatic ejection of fine dust.  None of these processes can offer a convincing explanation for  the high speeds reached by the ejecta in P/Vales.   In rotational disruption, for example, released material escapes with approximately the equatorial velocity of the rotation, typically $\le$ 1 m s$^{-1}$ at breakup for a kilometer-sized body.  Thermal fracture,  with realistic efficiencies for conversion of strain energy into kinetic energy of ejected fragments in the 1\% to 10\% range, produces peak speeds  $\sim$ (1 to 5) m s$^{-1}$ (Equation 14 of Jewitt 2012).  Electrostatic forces are weak, eject particles at similarly low speeds, and  cannot launch particles greater than micron-sized.  The mismatch with the ejecta speeds measured in P/Vales effectively  eliminates rotational disruption, thermal stresses and electrostatics as relevant mechanisms.

\clearpage

\section{SUMMARY}
We present time-resolved  observations of the 2010 photometric outburst of  P/2010 H2 (Vales). This is a quasi-Hilda  object, probably emplaced near the 3:2 mean-motion resonance  following a series of close encounters with Jupiter in the previous century.

\begin{enumerate}

\item The outburst, by $\Delta m_V\ge$ 7.5 magnitudes, started within a few hours of UT 2010 April 15.76, when 37 days past perihelion and at heliocentric distance 3.112 AU. 

\item Ejected particles had a maximum cross-section 17,600 km$^2$ and mass $\sim 1.2\times10^9$ kg (10$^{-4}$ of the nucleus mass), with radii from microns to centimeters following a differential size distribution $n(a)da \propto a^{-3.61\pm0.06}$. Peak ejection rates were 13,600 kg s$^{-1}$, and  the ejection appears to have been impulsive, with a timescale $\lesssim$1 day.

\item High measured   particle ejection speeds (up to 210 m s$^{-1}$) are compatible with gas drag acceleration from sublimated ice.
They are incompatible with  rotational instability, thermal and desiccation stress fracture, and electrostatic repulsion, all of which are therefore ruled out as driving mechanisms.   Asteroid impact, while  capable of generating high-speed ejecta, is improbable given the small population and short dynamical lifetime of the quasi-Hildas.

\item P/Vales is most likely a temporarily captured comet in which conductive heating of   sub-surface ice has triggered an outburst, perhaps through  exothermic crystallization from the  amorphous state.  If so, an ice volume about 0.4 km$^2$ in areal extent (about 1\% of the nucleus surface) and $\ge$5 m thick, and buried beneath a refractory layer a few meters thick, is inferred.  
\end{enumerate}

\acknowledgments
We thank Scott Sheppard for data from Magellan and comments on the manuscript, and Eric Hintz and Michael Joner  for data from the West Mountain Observatory.  We also thank Bin Yang and the two anonymous referees for comments. Some of the data presented herein were obtained at the W. M. Keck Observatory, which is operated as a scientific partnership among the California Institute of Technology, the University of California and NASA. The Observatory was made possible by the generous financial support of the W. M. Keck Foundation.  Y.K. was supported by the European Research Council (ERC) Starting Grant No. 757390 (CAstRA).



{\it Facilities:}  \facility{Magellan Observatory, Keck Observatory, West Mountain Observatory}.

\clearpage


\clearpage

\begin{deluxetable}{llcrrrccccr}
\tabletypesize{\scriptsize}
\tablecaption{P/Vales Timeline
\label{geometry}}
\tablewidth{0pt}
\tablehead{ \colhead{UT Date} & \colhead{Event/Telescope\tablenotemark{a}}  & \colhead{DOY\tablenotemark{b}} & \colhead{$r_H$\tablenotemark{c}} & \colhead{$\Delta$\tablenotemark{d}} & \colhead{$\alpha$\tablenotemark{e}}    & \colhead{$\nu$\tablenotemark{f}} & \colhead{$-\theta_{\odot}$\tablenotemark{g}} & \colhead{-$\theta_V$\tablenotemark{h}} & \colhead{$\delta_{\oplus}$\tablenotemark{i}}}
\startdata

2010 Mar 9.2  & Perihelion 		& 68.2 	& 3.108 & 2.291 & 12.1 & 0.0 & 273.2 & 303.3 & -6.0 \\
2010 April 15.76 & Outburst 		& 105.7 & 3.112 & 2.131 & 4.6 & 7.3 & 186.2 & 302.2 & -4.2 \\
2010 Apr 20.2 & Magellan 6.5 m 	& 110.2  & 3.113 & 2.137 & 5.3 & 8.2 & 167.8 & 302.1 & -3.8  \\
2010 Jun 6.3  &  WMO 0.9m 	&  157.3 & 3.131 & 2.489 & 16.1 & 17.4 & 118.2 & 302.2 & 1.1 \\
2010 Jun 15.3  &  WMO 0.9m 	& 166.3 & 	3.136 & 2.598 & 17.4 & 19.1 & 112.9 & 302.4 & 1.8 \\
2010 Jun 18.3  &  WMO 0.9m 	& 169.3 & 	3.137 & 2.640 & 17.7 & 19.7 & 115.5 & 302.5 & 2.0 \\
2010 Jul 3.2  &  WMO 0.9m 	& 184.2 &  	3.147 & 2.840 & 18.7 & 22.6 & 113.1 & 302.9 & 3.0 \\
2010 Aug 10.1  &  Keck 10 m 	&  222.1 & 	3.176 & 3.373 & 17.5 & 29.8 & 109.2 & 303.8 & 4.0 \\

\enddata


\tablenotetext{a}{Event or Telescope name and diameter}
\tablenotetext{b}{Day of Year, 1 = UT 2010 January 1}
\tablenotetext{c}{Heliocentric distance, in AU}
\tablenotetext{d}{Geocentric distance, in AU}
\tablenotetext{e}{Phase angle, degree}
\tablenotetext{f}{True Anomaly, degree}
\tablenotetext{g}{Position angle of the projected anti-solar vector, degree}
\tablenotetext{h}{Position angle of the projected negative velocity vector, degree}
\tablenotetext{i}{Angle of the Earth above the orbital plane, degree}
\end{deluxetable}

\clearpage

\begin{deluxetable}{cllll}
\tabletypesize{\scriptsize}
\tablecaption{Fixed Aperture R-band Photometry\tablenotemark{a}
\label{photometry}}
\tablewidth{0pt}
\tablehead{ \colhead{UT Date} &  \colhead{5,000 km}&  \colhead{10,000 km} & \colhead{20,000 km}  & \colhead{40,000 km} }
\startdata
2010 Apr 20.2 & 13.09$\pm$0.02/8.87/4.3 & 12.35$\pm$0.01/8.13/8.4 & 11.86$\pm$0.01/7.64/13.2 & 11.55$\pm$0.01/7.33/17.6 \\
2010 Jun 06.3 & 17.56$\pm$0.01/12.78/0.12 & 16.57$\pm$0.01/11.79/0.29 & 15.82$\pm$0.01/11.04/0.58 & 15.18$\pm$0.01/10.40/1.04 \\
2010 Jun 15.3 & 18.04$\pm$0.02/13.14/0.08 & 16.97$\pm$0.01/12.07/0.22 & 16.16$\pm$0.01/11.26/0.47 & 15.52$\pm$0.01/10.62/0.85 \\
2010 Jun 18.3 & 18.28$\pm$0.02/13.34/0.07 & 17.71$\pm$0.01/12.26/0.19 & 16.39$\pm$0.02/11.45/0.40 & 15.76$\pm$0.02/10.82/0.71 \\
2010 Jul 03.2 & 18.67$\pm$0.03/13.54/0.06 & 17.57$\pm$0.02/12.44/0.16 & 16.66$\pm$0.01/11.53/0.37 & 15.99$\pm$0.01/10.86/0.68 \\
2010 Aug 10.1\tablenotemark{b} & -- & -- & -- & 17.08$\pm$0.03/11.58/0.35\\
\enddata

\tablenotetext{a}{For each of four apertures of fixed projected radii 5$\times10^3$ km, 10$^4$ km, 2$\times10^4$ km and 4$\times10^4$ km we list the apparent red magnitude, m$_R$, the absolute R magnitude, $m_R$(1,1,0) and the scattering cross-section, $C_e$ in units of 10$^3$ km$^2$ in the format m$_R$/m$_R$(1,1,0)/$C_e$.}
\tablenotetext{b}{Inner apertures  omitted owing to the influence of very poor seeing on this date.}

\end{deluxetable}

\clearpage

\begin{deluxetable}{clcccccclll}
\tablecaption{Measured Optical Colors\tablenotemark{a}
\label{colors}}
\tablewidth{0pt}
\tablehead{\colhead{Color} &  \colhead{5,000 km} & \colhead{10,000 km}  & \colhead{20,000 km} & \colhead{40,000 km} & \colhead{Solar\tablenotemark{b}}}
\startdata
B-V & 0.89$\pm$0.02 & 0.90$\pm$0.02 & 0.85$\pm$0.02 & 0.74$\pm$0.02 & 0.64$\pm$0.02\\
V-R & 0.51$\pm$0.01 & 0.50$\pm$0.01 & 0.50$\pm$0.01 & 0.46$\pm$0.01 & 0.35$\pm$0.01 \\
R-I & 0.56$\pm$0.01 & 0.55$\pm$0.01 & 0.55$\pm$0.01 & 0.51$\pm$0.01 & 0.33$\pm$0.01\\

\hline

\enddata


\tablenotetext{a}{Data from UT 2010 April 20.2}
\tablenotetext{b}{Solar colors from Holmberg et al.~(2006)}

\end{deluxetable}

\clearpage
\clearpage

\clearpage

\clearpage

\clearpage

\clearpage

\clearpage

\clearpage

\clearpage

\clearpage

\clearpage

\end{document}